\documentclass[english,aps,prb,twocolumn,nofootinbib,showpacs,showkeys,secnumarabic,superscriptaddress,floatfix]{revtex4}

\usepackage[english]{babel}

\usepackage{amssymb,amsmath} 

\usepackage[pdftex]{graphicx}
\usepackage[caption=false,listofformat=subsimple]{subfig}

\usepackage{color} 
\usepackage[usenames,dvipsnames]{xcolor}
\usepackage[colorlinks=true,linkcolor=RoyalBlue,citecolor=OliveGreen,urlcolor=OliveGreen,linktoc=page]{hyperref} 

\usepackage{enumerate}
\usepackage{grffile}
\usepackage{balance}
\usepackage{float}

\newcommand{\Jc}{J_{\rm\scriptscriptstyle C}}
\newcommand{\Jdp}{J_{\rm dp}}

\newcommand{\Ec}{E_{\rm\scriptscriptstyle C}}

\newcommand{\Tc}{T_{\rm\scriptscriptstyle C}}
\newcommand{\Tcs}{T_{\rm\scriptscriptstyle C,S}}
\newcommand{\Tci}{T_{\rm\scriptscriptstyle C,I}}
\newcommand{\es}{\varepsilon_{\rm\scriptscriptstyle S}}
\newcommand{\ei}{\varepsilon_{\rm\scriptscriptstyle I}}

\newcommand{\Hct}{H_{\rm{\scriptscriptstyle C}2}}
\newcommand{\Bm}{B_{\scriptscriptstyle\Phi}} 
\newcommand{\Bmn}{B_{\scriptscriptstyle\Phi, \rm N}} 
\newcommand{\Bmc}{B_{\scriptscriptstyle\Phi, \rm C}} 
\newcommand{\Bmx}{B_{\scriptscriptstyle\Phi, \rm X}} 

\newcommand{\fN}{f_{\rm\scriptscriptstyle N}}
\newcommand{\fC}{f_{\rm\scriptscriptstyle C}}
\newcommand{\fX}{f_{\rm\scriptscriptstyle X}}
\newcommand{\fP}{f_{\rm\scriptscriptstyle P}}

\begin{document}

\title{Toward superconducting critical current by design}

\author{Ivan A. Sadovskyy} 
\affiliation{Materials Science Division, Argonne National Laboratory, Argonne, Illinois 60439}

\author{Ying Jia}
\affiliation{Materials Science Division, Argonne National Laboratory, Argonne, Illinois 60439}

\author{Maxime Leroux}
\affiliation{Materials Science Division, Argonne National Laboratory, Argonne, Illinois 60439}

\author{Jihwan Kwon}
\affiliation{Department of Materials Science and Engineering, University of Illinois-Urbana Champaign, Urbana, Illinois 61801}

\author{Hefei Hu}
\affiliation{Materials Research Laboratory, University of Illinois-Urbana Champaign, Urbana, Illinois 61801}

\author{Lei Fang}
\affiliation{Materials Science Division, Argonne National Laboratory, Argonne, Illinois 60439}
\affiliation{Department of Chemistry, Northwestern University, Evanston, Illinois 60208}

\author{Carlos Chaparro}
\affiliation{Materials Science Division, Argonne National Laboratory, Argonne, Illinois 60439}

\author{Shaofei Zhu}
\affiliation{Physics Division, Argonne National Laboratory, Argonne, Illinois 60439}

\author{Ulrich Welp}
\affiliation{Materials Science Division, Argonne National Laboratory, Argonne, Illinois 60439}

\author{Jian-Min Zuo}
\affiliation{Department of Materials Science and Engineering, University of Illinois-Urbana Champaign, Urbana, Illinois 61801}
\affiliation{Materials Research Laboratory, University of Illinois-Urbana Champaign, Urbana, Illinois 61801}

\author{Yifei Zhang}
\affiliation{SuperPower Corp., Schenectady, New York 12304}

\author{Ryusuke Nakasaki}
\affiliation{SuperPower Corp., Schenectady, New York 12304}

\author{Venkat Selvamanickam}
\affiliation{Department of Mechanical Engineering and Texas Center for Superconductivity, University of Houston, Houston, Texas 77204}

\author{George W. Crabtree}
\affiliation{Materials Science Division, Argonne National Laboratory, Argonne, Illinois 60439}
\affiliation{University of Illinois at Chicago, Chicago, Illinois 60607}

\author{Alexei E. Koshelev}
\affiliation{Materials Science Division, Argonne National Laboratory, Argonne, Illinois 60439}

\author{Andreas Glatz}
\affiliation{Materials Science Division, Argonne National Laboratory, Argonne, Illinois 60439}
\affiliation{Department of Physics, Northern Illinois University, DeKalb, Illinois 60115}

\author{Wai-Kwong Kwok}
\affiliation{Materials Science Division, Argonne National Laboratory, Argonne, Illinois 60439}

\date{March 31, 2016}

\begin{abstract}
We present the new paradigm of critical current by design. Analogous to materials by design, it aims at predicting the optimal defect landscape in a superconductor for targeted applications by elucidating the vortex dynamics responsible for the bulk critical current. To highlight this approach, we demonstrate the synergistic combination of critical current measurements on commercial high-temperature superconductors containing self-assembled and irradiation tailored correlated defects by using large-scale time-dependent Ginzburg-Landau simulations for vortex dynamics.
\end{abstract}



\maketitle

\section{Introduction}

The interaction of vortex matter with defects in applied superconductors directly determines their current carrying capacity. Defects range from chemically grown nanostructures and crystalline imperfections to the layered structure of the material itself. The vortex-defect interactions are non-additive in general, leading to complex dynamic behavior that has proven difficult to capture in analytical models. With recent progress in advanced computing, a new paradigm has emerged that aims at simulation-assisted design of defect structures with predictable powers. This concept of \textit{critical current by design} aims at predicting the optimal defect landscape for targeted applications by elucidating the vortex dynamics responsible for the bulk critical current. To highlight this approach, we demonstrate the synergistic combination of critical current measurements on commercial high-temperature superconductors containing self-assembled and irradiation tailored correlated defects with large-scale time-dependent Ginzburg-Landau numerical simulations of vortex dynamics. The qualitative agreement between experiments and simulations allows making numerical predictions for a wider parameter range and hence opens the route to the critical-current-by-design paradigm.

Superconducting vortices appear inside superconductors in the presence of sufficiently large magnetic fields and are responsible for the entire electromagnetic behavior of applied superconducting materials. In particular, the onset of vortex motion limits the highest dissipation-less electrical current~--- the critical current~--- the superconductor can carry. At the microscopic level, vortices can be viewed as non-superconducting cores that are surrounded by circulating persistent superconducting currents. The radius of the core is approximately given by the superconducting coherence length, $\xi$, which for high-temperature superconductors is of the order of 1.5--3\,nm, and the circulating supercurrents extend out to the magnetic penetration depth, $\lambda$, which is of the order of 100--200\,nm. When an external electric current is passed through the superconductor, the vortices experience the Lorentz force, which sets them in motion, resulting in a voltage drop across the superconductor and hence dissipation: the superconductor looses its superconducting properties.\cite{Campbell:1972} A variety of defects in the superconducting material (precipitates, point defects, grain boundaries, dislocations, stacking faults, strain fields, etc.) can induce spatial variations of the superconducting state that serve to pin the normal vortex cores and prevent their motion. Thus, an important goal of materials science of applied superconductors is the design and synthesis of the most effective vortex pinning microstructures that results in the largest critical current. Although a tremendous amount of knowledge, systematics and practical experience has been gained in this pursuit, see for instance reviews,\cite{Kramer:1973,Dew-Hughes:1974,Kes:1992,Scanlan:2004,Godeke:2006} the fundamental solution to the dynamics of interacting elastic vortex lines in a disordered medium is still an open problem. The technical challenge consists in the accurate statistical summation of individual vortex-vortex interactions and pinning forces \cite{Campbell:1972,Larkin:1979} which is akin to a many-body problem with long-range interactions. For example, the critical currents may be estimated based on energy and force balance considerations, see, e.g., reviews \onlinecite{Blatter:1994,Blatter:2003,Brandt:1995} only for the most simple cases, such as uniformly distributed weak pinning centers. These estimates, however, do not describe the more practically important cases of multiple large-size defects occupying a considerable fraction of the material, as in most commercial high temperature superconductors. In such cases, the vortex pinning effects of different pinning sites are not simply additive, especially for high concentration of defects and in the presence of strong magnetic fields. In general, critical currents for such complicated defect structures can only be computed using large-scale numerical modeling.

The rapid progress of high performance computing combined with sophisticated numerical methods and materials synthesis provides a strong impetus to realize approaches with predictive power. A new area driven by this powerful combination is the materials genome initiative\cite{Curtarolo:2013,MGI} that aims at predicting materials with desired properties: \textit{materials by design}. An analogous \textit{critical-current-by-design} concept can be envisioned in applied superconductivity, with the use of large-scale numerical simulations for the rational design of defect microstructures in order to optimize the current-carrying capacity of superconducting materials for targeted applications. The building blocks for this design are materials defects, which range from atomic to mesoscopic in size and their interaction with superconducting vortices.

\section{Model}

A well-established description of superconducting properties is based on the phenomenological time-dependent Ginzburg-Landau (TDGL) model.\cite{Schmid:1966,Aranson:2002} In this model, the superconducting state is represented by a complex-valued superconducting order parameter depending on time and coordinate, $\psi(\textbf{r}, t)$. Vortices enter as topological defects associated with a phase-winding of $2\pi$ in the order parameter field. This model automatically accounts for the interactions between vortices, between vortices and pin sites, for vortex-line elasticity, vortex cutting and re-connection, in other words, all the interactions that are relevant for describing pinning. The TDGL model can be reduced to a non-linear partial differential equation for the order parameter, given here in dimensionless form
\begin{equation}
	(\partial_t + i\mu)\psi 
	= \varepsilon(\textbf{r}) \psi - |\psi|^2 \psi 
	+ (\nabla - i\textbf{A})^2 \psi + \zeta(\textbf{r}, t).
	\label{eq:GL}
\end{equation} 
We use the temperature-dependent coherence length $\xi = \xi(T)$ as unit of length, which allows to account for the temperature dependence of physical quantities in a computationally efficient way. The micro and nanostructure of the superconductor is modeled through the spatial dependence of the critical temperature $\Tc(\textbf{r})$ contained in the prefactor of the linear term, $\varepsilon(\textbf{r})$. This prefactor is $\varepsilon(\textbf{r}) = 1$ in the superconductor and is negative, $\varepsilon(\textbf{r}) < 0$, in the non-superconducting defect. $\textbf{A}$ is the vector potential describing the magnetic field as $\textbf{B} = \mathrm{curl}\textbf{A}$, $\mu$ is the electrostatic potential, and $\zeta(\textbf{r}, t)$ is the temperature-dependent Langevin noise. We employed the so-called infinite-$\lambda$ approximation within which the vector potential is fixed by the applied magnetic field. This approximation is suitable for type-II superconductors with high Ginzburg-Landau parameter ($\lambda/\xi \sim 100$ for the experimental system) at high magnetic fields, when the intervortex distance is much smaller than the London penetration depth~$\lambda$ and small spatial variations of the magnetic field can be neglected. The details of the numerical simulations are presented in Appendix~\ref{sec:numerical_simulations}.

For realistic pinning microstructures, Equation~\eqref{eq:GL} can be solved only numerically. This task proves to be computationally very demanding, and thus far, mostly mesoscopic or two-dimensional systems have been analyzed with this method, see e.g. Refs.~\onlinecite{Schweigert:1998,Crabtree:2000,Berdiyorov:2006,Vodolazov:2013}. Recently, we have developed a stable implicit iterative solver for TDGL equation\cite{Sadovskyy:2015a} based on the Jacobi method. This solver running on high-performance general-purpose graphic processor units allows to simulate three-dimensional samples large enough that finite size and surface effects become negligible.\cite{Koshelev:2015,Sadovskyy:2016} Predictions for macroscopic quantities such as the critical current are now possible, thereby enabling the concept of critical current by design. We demonstrate this concept on technologically important rare earth barium copper oxide (REBa$_2$Cu$_3$O$_{7-\delta}$ or \mbox{REBCO}) coated conductors and validate the results on samples with clearly identifiable pinning effects due to linear correlated defects in the form of self-assembled barium zirconate (BZO) nanorods and irradiation tracks introduced by heavy-ion irradiation.

\begin{figure*}
	\begin{center}
		\subfloat{\includegraphics[width=11.9cm]{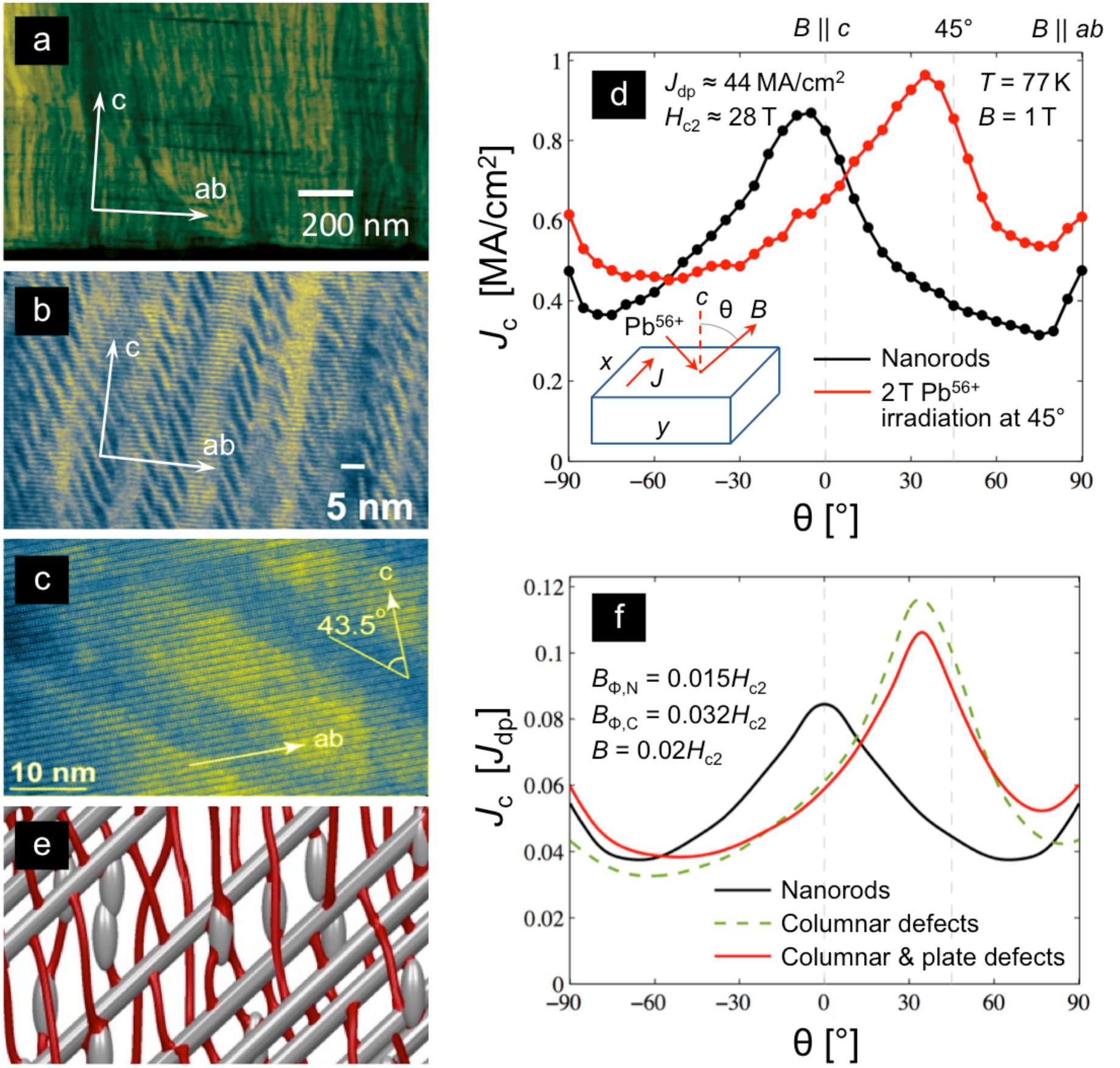} \label{fig:nanorods_TEM_200nm}}
		\subfloat{\label{fig:nanorods_TEM_5nm}}
		\subfloat{\label{fig:nanorods_irradiated_TEM}}
		\subfloat{\label{fig:nanorods_irradiated_Jc_experiment}}
		\subfloat{\label{fig:nanorods_irradiated_order_parameter}}
		\subfloat{\label{fig:nanorods_irradiated_Jc_theory}}
	\end{center} \vspace{-5mm}
	\caption{
		\protect\subref{fig:nanorods_TEM_200nm}.~Transmission electron microscopy (TEM) 
		image of the unirradiated sample with self-assembled nanorods aligned along the $c$-axis. 
		\protect\subref{fig:nanorods_TEM_5nm}.~High-resolution TEM image of BZO nanorods. 
		The nanorods are oriented along the $c$-axis, and the dark striations are Moire-patterns 
		resulting from the imaging setup. 
		\protect\subref{fig:nanorods_irradiated_TEM}.~Heavy-ion damage tracks revealed 
		by Z-contrast scanning TEM in a cross-section view along the [100] direction 
		with atomic resolution. The tracks are inclined at $\sim 45^\circ$ from the $c$-axis. 
		In panels~\protect\subref{fig:nanorods_TEM_200nm}--\protect\subref{fig:nanorods_irradiated_TEM} 
		the color scheme was chosen to enhance the contrast. 
		\protect\subref{fig:nanorods_irradiated_Jc_experiment}.~Angular dependence of the critical 
		current $\Jc(\theta)$ before and after irradiation in REBCO coated conductor samples 
		at a temperature of 77\,K and in a magnetic field of $B = 1$\,T. The sample was irradiated 
		with 1.4\,GeV Pb$^{56+}$ ions at $45^\circ$ from the $c$-axis. In the unirradiated sample, 
		the presence of BZO nanorods induces a large peak when the field is applied parallel 
		to the $c$-axis of the REBCO layer, and a small peak is also present for the $ab$-plane 
		due to intrinsic pinning, RE$_2$O$_3$ nanoplates, and stacking faults. In the irradiated 
		sample, the peak has shifted toward the direction of irradiation. 
		\protect\subref{fig:nanorods_irradiated_order_parameter}.~Representative configuration 
		of vortices and defects in the simulation. A snapshot of the isosurfaces of the order 
		parameter is shown in red. The vortices follow the direction of the magnetic field applied 
		along the $c$-axis. Columnar defects and nanorods (displayed shorter than actually 
		used in the simulations for illustration purposes) are shown in grey. See the videos 
		of vortex motion in the sample with short nanorods (magnetic filed $B$ 
		applied \href{http://youtu.be/3e7IvkHGwb8}{along the $c$-axis} and 
		\href{http://youtu.be/kM1nBS7BFjY}{at $45^\circ$ to the $c$-axis}) and 
		in the sample with nanorods and tilted columnar defects 
		($B$ \href{http://youtu.be/Jqtu8xI7qYY}{along the $c$-axis} and 
		\href{http://youtu.be/FHboiE0fl8Q}{at $45^\circ$ to the $c$-axis}). 
		\protect\subref{fig:nanorods_irradiated_Jc_theory}.~Angular dependence 
		of the critical current from simulations, normalized to the depairing current. 
		Reference simulation with sample containing nanorods along the $c$-axis is shown in black. 
		The addition of columnar defects at 45$^\circ$ from the $c$-axis shifts the peak as shown 
		by the green dotted line. The addition of plate-like defects in the $ab$-plane 
		modifies the angular dependence as shown in red.
	}
	\label{fig:nanorods_irradiated}
\end{figure*}

\section{Results and discussion}

We present results on two sets of commercially available REBCO coated conductors, one containing self-assembled BZO nanorods, the other without. The samples were grown on Hastelloy substrates using metal organic chemical vapor deposition (MOCVD). The in-plane texture of a MgO buffer layer located between the superconductor and the substrate is achieved with an ion beam assisted deposition (IBAD) process.\cite{Selvamanickam:2009} It has recently been shown that the addition of BZO can lead to the formation of self-assembled BZO nanorods inside the REBCO matrix during the synthesis of the superconductor.\cite{MacManus-Driscoll:2004,Haugan:2004,Kang:2006,Maiorov:2009} The nanorods grow largely along the $c$-axis of REBCO with typical diameters of $\sim 5$--10\,nm and lengths of several hundreds of nanometers that is mostly limited by planar precipitates and stacking faults. Their addition results in substantial increases of the critical current density with $\Jc$ values reaching $\approx 8$\,MA/cm$^2$ at 30\,K and in a field of 9\,T applied parallel to $c$-axis.\cite{Selvamanickam:2015} At high temperatures ($T \gtrsim 60$\,K), the linear shape of the BZO nanorods is responsible for the pronounced peak in the field-angle dependence of $\Jc$ for applied fields aligned with the $c$-axis, whereas at lower temperatures, strain fields associated with the BZO nanorods and/or nanoparticles induce almost isotropic $\Jc$-enhancements.\cite{Selvamanickam:2015,Xu:2012}

An alternative approach to introducing linear defects is realized by irradiation with high-energy heavy ions.\cite{Toulemonde:1994} For certain ranges of parameters~--- such as weight and energy of the ions, thermal and electric conductivity of the target material, and rate of energy transfer, continuous amorphous tracks are formed in the target material. Typical track diameters fall in the range of 5--10\,nm.\cite{Toulemonde:1994,Wheeler:1993,Zhu:1993} These heavy-ion induced columnar defects are known to be the most effective vortex pinning centers for applied fields aligned with the tracks due to their ideal size for confining vortices, and their pinning effects have been extensively studied.\cite{Civale:1991,Civale:1997,Fang:2013} In contrast to self-assembled BZO nanorods that are chemically inserted, irradiation induced columnar defects can be introduced at arbitrary angles and over a wide range of areal density,\cite{Krusin-Elbaum:1996a,Krusin-Elbaum:1996b,Kwok:1998a,Kwok:1998b} allowing for a convenient means of investigating the constructive or destructive pinning action of several types of coexisting pinning sites.

In this work, we present (i)~a study of vortex pinning in a REBCO coated conductor with self-assembled BZO nanorods before and after heavy-ion irradiation at 45$^\circ$ and (ii)~a study of the cumulative effect of splayed heavy ion irradiation at angles of $+45^\circ$ and $-45^\circ$ to the $c$-axis in a REBCO coated conductor without nanorods. In both studies, we compare experimental and numerical results and demonstrate the validity of our numerical TDGL approach. Subsequently we explore a wider parameter range with numerical simulations to determine optimal pinning configurations.

\subsection{Sample with nanorods irradiated at an angle}

\begin{figure*}
	\begin{center}
		\subfloat{\includegraphics[width=14cm]{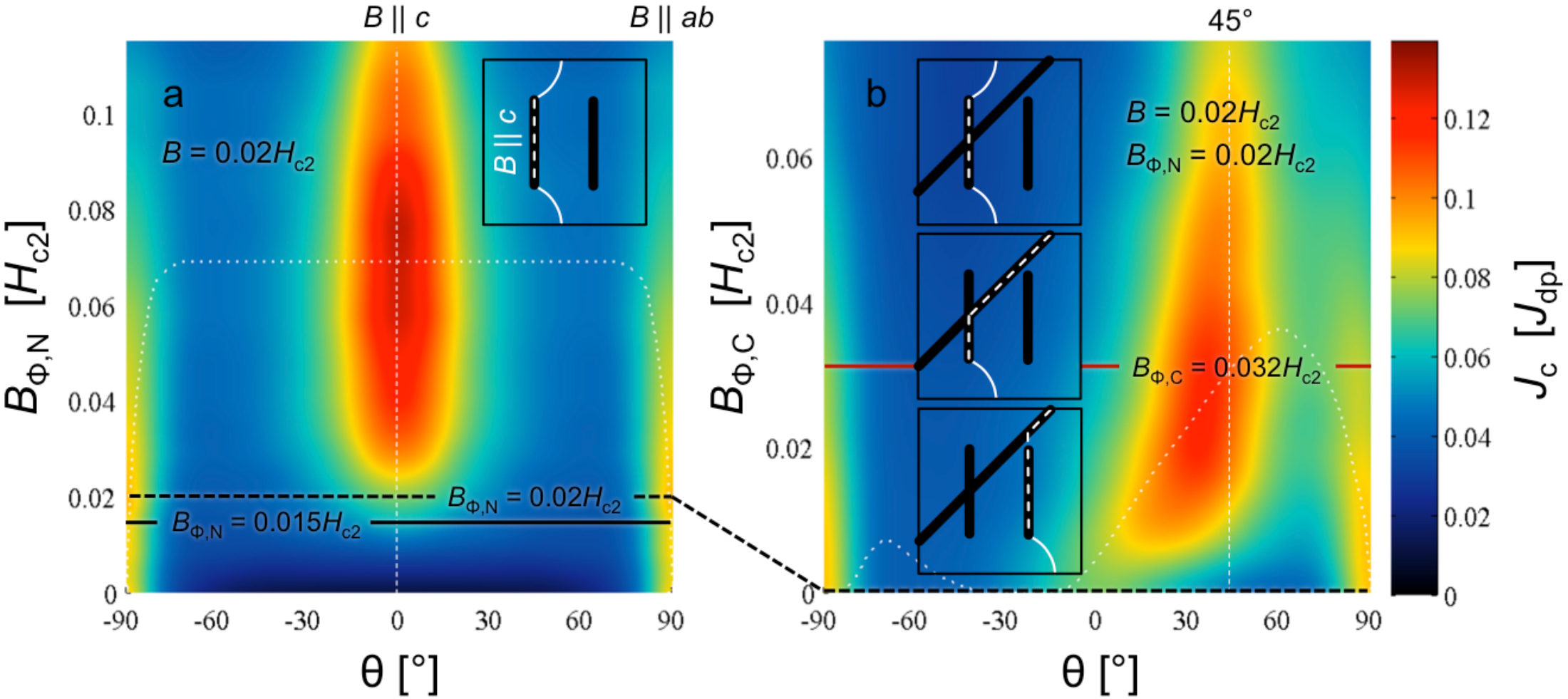} \label{fig:nanorods_Jc_concentration_theory}}
		\subfloat{\label{fig:irradiated_Jc_concentration_theory}}
	\end{center} \vspace{-5mm}
	\caption{
		\protect\subref{fig:nanorods_Jc_concentration_theory}.~Angular dependence 
		of the critical current (normalized to the depairing current $\Jdp$) for different 
		concentrations of nanorods in a magnetic field of $0.02\Hct$. One can see 
		the prominent peak along the $c$-axis ($\theta = 0^\circ$) for a wide range 
		of nanorod concentration (expressed in matching field $\Bmn$) ranging 
		from $\Bmn \approx 0.025\Hct$ to $\approx 0.1\Hct$. This corresponds 
		to the volume fraction $\approx 5$--20\% occupied by nanorods inclusions. 
		The peak critical current at $\theta = 0^\circ$ and $\Bmn \approx 0.07\Hct$ 
		is $\Jc = 0.14\Jdp$. The white dotted curve shows the optimal concentration 
		of nanorods for maximum critical current at a given angle of applied field. 
		Inset: Sketch of the vortex pinned by a nanorod for $B||c$. 
		\protect\subref{fig:irradiated_Jc_concentration_theory}.~Angular dependence 
		of the critical current for different concentration of columnar defects $\Bmc$ 
		tilted at $\theta = 45^\circ$ and at a fixed concentration of nanorods 
		(with matching field $\Bmn \approx 0.02\Hct$ shown by dashed line 
		in panel~\protect\subref{fig:nanorods_Jc_concentration_theory}). 
		With increasing concentration of columnar defects, the peak shifts 
		from $\theta = 0^\circ$ to 45$^\circ$. The near-optimal critical current corresponds 
		to the angular range $\theta = 30$--40$^\circ$ and concentration of columnar 
		defects $\Bmc \approx 0.02$--$0.04\Hct$ 
		(or volume fraction $\approx 6$--11\% occupied by them). 
		Inset: Vortex sliding on tilted columnar defects.
	}
	\label{fig:nanorods_irradiated_Jc_concentration_theory}
\end{figure*}

Figure~\ref{fig:nanorods_irradiated} summarizes our results on a REBCO coated conductor containing BZO nanorods into which columnar-shaped tracks have been introduced at an angle close to 45$^\circ$ from the $c$-axis using 1.4\,GeV $^{208}$Pb ion irradiation. The unirradiated sample (Figures~\ref{fig:nanorods_TEM_200nm} and \ref{fig:nanorods_TEM_5nm}) contains BZO nanorods with estimated density corresponding to matching field $\Bmn \sim 1$\,T (meaning the effective defect density is equivalent to the vortex density at a field of 1\,T). After irradiation (Figure~\ref{fig:nanorods_irradiated_TEM}) sample contains $9.7 \times 10^{10}$ tracks/cm$^2$ corresponding to matching field of $\Bmc \approx 2$\,T. (We distinguish the matching field for irradiated columnar defects, $\Bmc$, from that of nanorods, $\Bmn$, and from for the total density of splayed columnar defects irradiated in two directions, $\Bmx$.) These irradiation induced columnar defects have an amorphous core of $\sim 8$\,nm in diameter, similar to Pb-ion irradiation damage in YBCO single crystals and extend through the entire thickness of the film. Figure~\ref{fig:nanorods_irradiated_Jc_experiment} shows a comparison of the angular dependence of $\Jc$ from transport measurements in unirradiated (black) and irradiated at 45$^\circ$ (red) samples with self-assembled BZO nanorods. In these measurements, the current flows along the $ab$-plane and is always perpendicular to the applied magnetic field. The field is rotated from the $c$-axis ($\theta = 0^\circ$) to the $ab$-plane ($\theta = 90^\circ$). The irradiation-induced columnar tracks are oriented in such a way that they are perpendicular to the current flow.

\begin{figure*}
	\begin{center}
		\subfloat{\includegraphics[width=11.7cm]{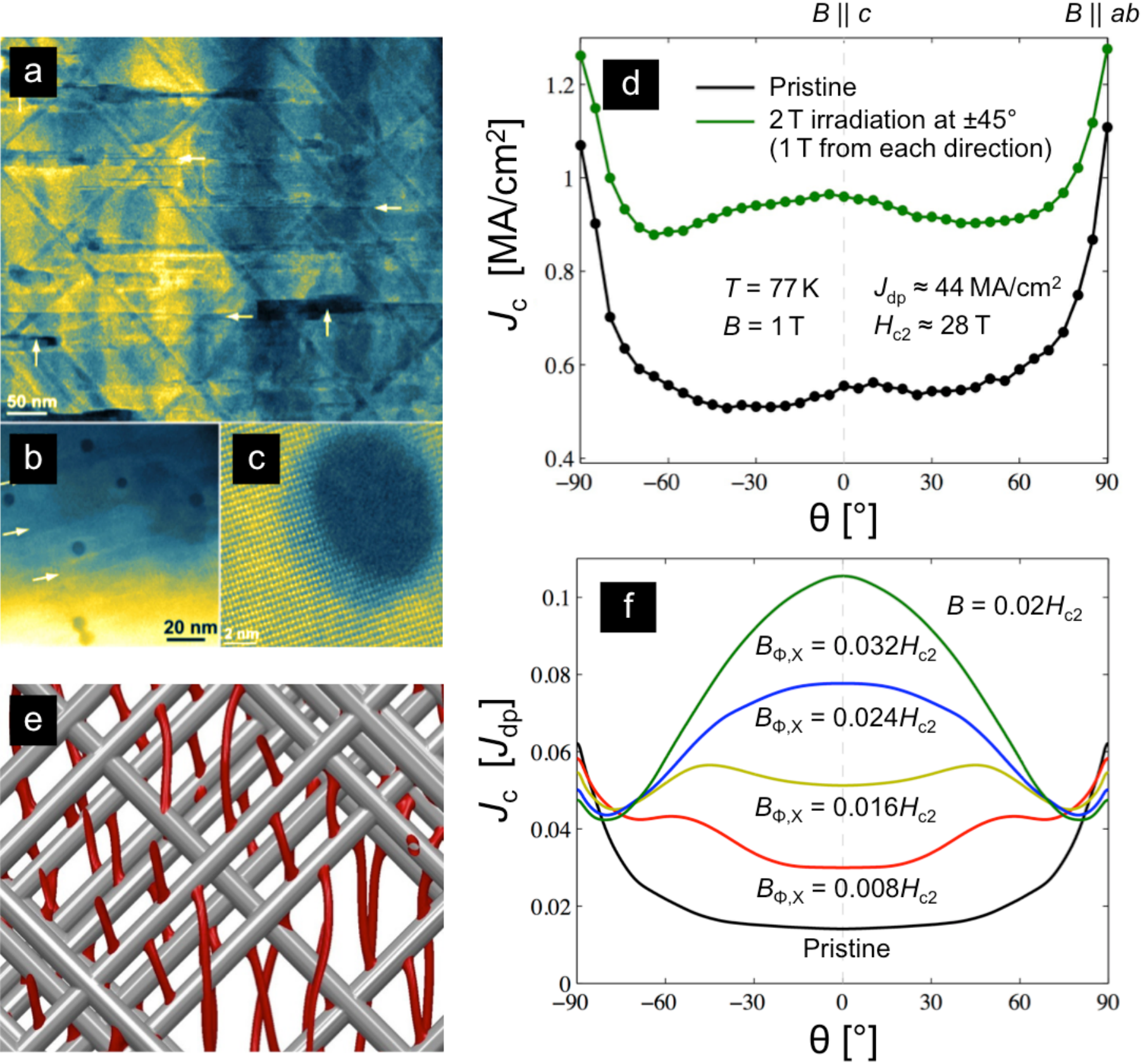} \label{fig:splayed_TEM_50nm}}
		\subfloat{\label{fig:splayed_TEM_20nm_top}}
		\subfloat{\label{fig:splayed_TEM_2nm_top}}
		\subfloat{\label{fig:splayed_Jc_experiment}}
		\subfloat{\label{fig:splayed_order_parameter}}
		\subfloat{\label{fig:splayed_Jc_theory}}
	\end{center} \vspace{-5mm}
	\caption{
		\protect\subref{fig:splayed_TEM_50nm}.~Z-contrast scanning TEM image 
		along the [100] direction of the $\Bmx = 2$\,T, $\pm 45^\circ$ irradiated 
		REBCO tape. The damage tracks are at $\pm 45^\circ$ from the $c$-axis 
		and extend through the entire thickness of the REBCO film. 
		The pre-existing stacking faults and nanoplates are indicated 
		by horizontal and vertical arrows respectively. 
		\protect\subref{fig:splayed_TEM_20nm_top}.~TEM image along 
		the $+45^\circ$ irradiation direction showing a cross-section with 
		the tracks appearing as black dots. 
		\protect\subref{fig:splayed_TEM_2nm_top}.~High magnification image 
		showing the cross-section of a damage track, $\approx 8$\,nm wide. 
		The featureless dark region indicates an amorphous structure. 
		The elliptic shape and the shadow are due to a small misalignment 
		between the irradiation direction and the sample-cutting direction. 
		In panels~\protect\subref{fig:splayed_TEM_50nm}--\protect\subref{fig:splayed_TEM_2nm_top} 
		the color scheme was chosen to enhance the contrast. 
		\protect\subref{fig:splayed_Jc_experiment}.~Angular dependence 
		of the critical current in pristine and irradiated (1\,T in each direction) 
		REBCO tape samples at 77\,K. The critical current performance 
		is significantly improved by the incorporation of splayed columnar defects. 
		\protect\subref{fig:splayed_order_parameter}.~Isosurfaces of the order parameter 
		in a model with splayed columnar defects and an external magnetic field applied 
		along the $c$-axis, also see the videos for magnetic filed applied 
		\href{http://youtu.be/z59VGyYfYLU}{along the $c$-axis} and 
		\href{http://youtu.be/IutqyRR0MKU}{at $45^\circ$ to the $c$-axis}.
		\protect\subref{fig:splayed_Jc_theory}.~Numerically simulated angular dependence 
		of the critical current for several densities $\Bmx$ of splayed columnar defects.
	}
	\label{fig:splayed}
\end{figure*}

The angular dependence of the critical current $\Jc(\theta)$ conveniently characterizes the nature of the dominant correlated pinning centers. Pinning contributions from the inherent layered structure of the material, from in-plane nanoplates and from stacking faults create two strong peaks in $\Jc$ for fields oriented along the $ab$-plane, at $\pm 90^\circ$. As expected, the sample with BZO nanorods presents a pronounced peak in $\Jc$ near $B||c$\cite{MacManus-Driscoll:2004,Maiorov:2009,Xu:2014} but at the expense of the in-plane critical current $\Jc(B||ab)$.\cite{Xu:2010,Polat:2012} As a result, at 77\,K and $B = 1$\,T, the BZO nanorods actually reverse the critical-current anisotropy from 2 to 0.5, defined as $\Jc(B||ab)/\Jc(B||c)$ as can be seen by comparing the data labeled `pristine' in Figure~\ref{fig:splayed_Jc_experiment} and the data labeled `nanorods' in Figure~\ref{fig:nanorods_irradiated_Jc_experiment}. 

Remarkably, in the sample containing both BZO nanorods and irradiation-induced columnar defects, we observe only a single peak in the critical current at $\theta \approx 45^\circ$, instead of the superposition of a $\Jc$-peak due to the BZO nanorods at $\theta \approx 0^\circ$ and a $\Jc$-peak due to the columnar defects at $\theta \approx 45^\circ$. Namely, the vortex pinning due to the BZO nanorods has been strongly suppressed by the introduction of the irradiation tracks. Such an effect is a clear example of the non-additive nature of two types of pinning centers. 

When describing this situation theoretically, we mainly focus on the two types of pinning centers that dominate the angular dependence of the critical current and of which we have quantitative knowledge: nanorods and irradiation induced columnar defects. Furthermore, we model intrinsic pinning by means of the Lawrence-Doniach model\cite{Lawrence:1970} and a modified Laplacian describing anisotropy. In addition, we consider contributions from in-plane plate-like defects. The model of the tape nanostructure, that includes both the BZO nanorods aligned along the $c$-axis and the columnar defects inclined at $\theta = 45^\circ$, is presented in Figure~\ref{fig:nanorods_irradiated_order_parameter}.

Our simulation allows for an unprecedented insight into the three-dimensional vortex dynamics as shown in Figures~\ref{fig:nanorods_irradiated_order_parameter} and \ref{fig:splayed_order_parameter}. We find that for large enough concentrations of irradiation-induced columnar defects, vortices aligned with the BZO nanorods can easily jump from one nanorod to the next by sliding along the oblique columnar defects. This effect is directly responsible for the reduced critical current for magnetic field along the BZO nanorod direction. The resulting simulated angular dependences of $\Jc$ are shown in Figure~\ref{fig:nanorods_irradiated_Jc_theory} and are in good qualitative agreement with the experiment. In particular, they depict the non-additive effect of the two types of pinning centers and that the pinning action of the BZO nanorods is no longer visible. We therefore attribute the loss of pinning by the nanorods at $\theta = 0^\circ$ observed in the experiment to a `vortex sliding' effect between nanorods induced by the oblique continuous columnar tracks, see insets in Figure~\ref{fig:nanorods_irradiated_Jc_concentration_theory}. In other words, the columnar defects tilted at $45^\circ$ effectively reduce the nucleation energy of local vortex kink formation and thus, help vortices to slide from one nanorod to the next one.

We typically observe that dominant extended defects define the critical current peak of $\Jc(\theta)$ at their tilt angle, while all other defects only slightly modify the shape of $\Jc(\theta)$, without adding additional peaks. For example, coated conductor samples contain a fair amount of stacking faults and $ab$-plane precipitates that disrupt the columnar defects and the BZO nanorods. To elucidate their role, we simulate the system with and without in-plane, plate-like defects. The predominant effect of these planar defects is an enhancement of the critical current for in-plane magnetic fields, $\theta = 90^\circ$. Comparing the dashed green and the red lines in Figure~\ref{fig:nanorods_irradiated_Jc_theory}, one sees that the in-plane defects slightly reduce the peak in fields aligned with the columnar defects, $\theta \approx 45^\circ$. 

Additionally, to obtain predictive results, we numerically analyze a wide range of concentrations of BZO nanorods and columnar defects. In Figure~\ref{fig:nanorods_irradiated_Jc_concentration_theory} we show density plots of the simulated critical current as a function of angle $\theta$ and concentration of the defects characterized by the matching field $\Bm$. Figure~\ref{fig:nanorods_Jc_concentration_theory} shows that with increasing nanorod concentration (without columnar defects) a prominent maximum in $\Jc(\theta)$ emerges around $\theta = 0^\circ$. For nanorod concentrations ranging from $\Bmn \approx 0.025\Hct$ to $\approx 0.1\Hct$ the height and angular width of this maximum are predicted to be almost independent of concentration. With further increase of the nanorod concentration, the $\Jc$ peak at $\theta = 0^\circ$ decreases. 

Figure~\ref{fig:nanorods_Jc_concentration_theory} also shows that the two $\Jc$ peaks at $\theta = \pm 90^\circ$ attributed to intrinsic pinning from the material's layered structure are suppressed when the nanorod defect density exceeds $\Bmn \approx 0.04\Hct$. The critical current for $B||c$ approaches its maximum at $\Bmn \approx 0.07\Hct$, which corresponds to a volume fraction of 14\% occupied by nanorods. We find that the mechanism of this suppression is the same as described above for vortices jumping between nanorods by `sliding' through the columnar defects, but this time it is the nanorods that create `bridges' on which vortices can move from one intrinsic layer to the next. Experimentally, we note that the critical current for $B||ab$ due to intrinsic pinning above an over-all isotropic baseline level indeed decreases with increasing concentration of BZO nanorods, in agreement with our simulations. This trend holds even though the baseline $\Jc$ and additional pinning from planar precipitates~--- both not included in the current simulations~--- vary significantly between samples. 

Figure~\ref{fig:irradiated_Jc_concentration_theory} shows the effect of increasing the concentration of irradiation induced columnar defects for a fixed concentration of BZO nanorods of $\Bmn \approx 0.02\Hct$. The intrinsic pinning peaks at $\pm 90^\circ$ vanish above a dose matching field of $\Bmc \approx 0.06\Hct$. A similar reduction of intrinsic pinning was observed experimentally in YBCO films that contain inclined columnar defects.\cite{Holzapfel:1993} Furthermore, the peak at $\theta = 0^\circ$ due to the BZO nanorods rapidly shifts to $\approx 45^\circ$ with increasing columnar defect concentration from $\Bmc \approx 0.02\Hct$ to $\approx 0.04\Hct$. For even larger columnar defect concentrations, this latter peak becomes smaller as the effective barrier for vortex hopping between neighboring columnar defects decreases. This shift in the $\Jc$ peak when BZO nanorods compete with columnar defects oriented at a different angle clearly highlights the complex vortex dynamics in a mixed pinning environment. Although newly added defects can pin vortices, they may also facilitate vortex hopping or sliding between neighboring preexisting defects under specific conditions. Hence the effect of these defects on vortices can be tailored by controlling the concentrations and mutual orientations of the defects.

\subsection{Pristine sample irradiated at two angles}

\begin{figure}
	\begin{center}
		\includegraphics[width=7.75cm]{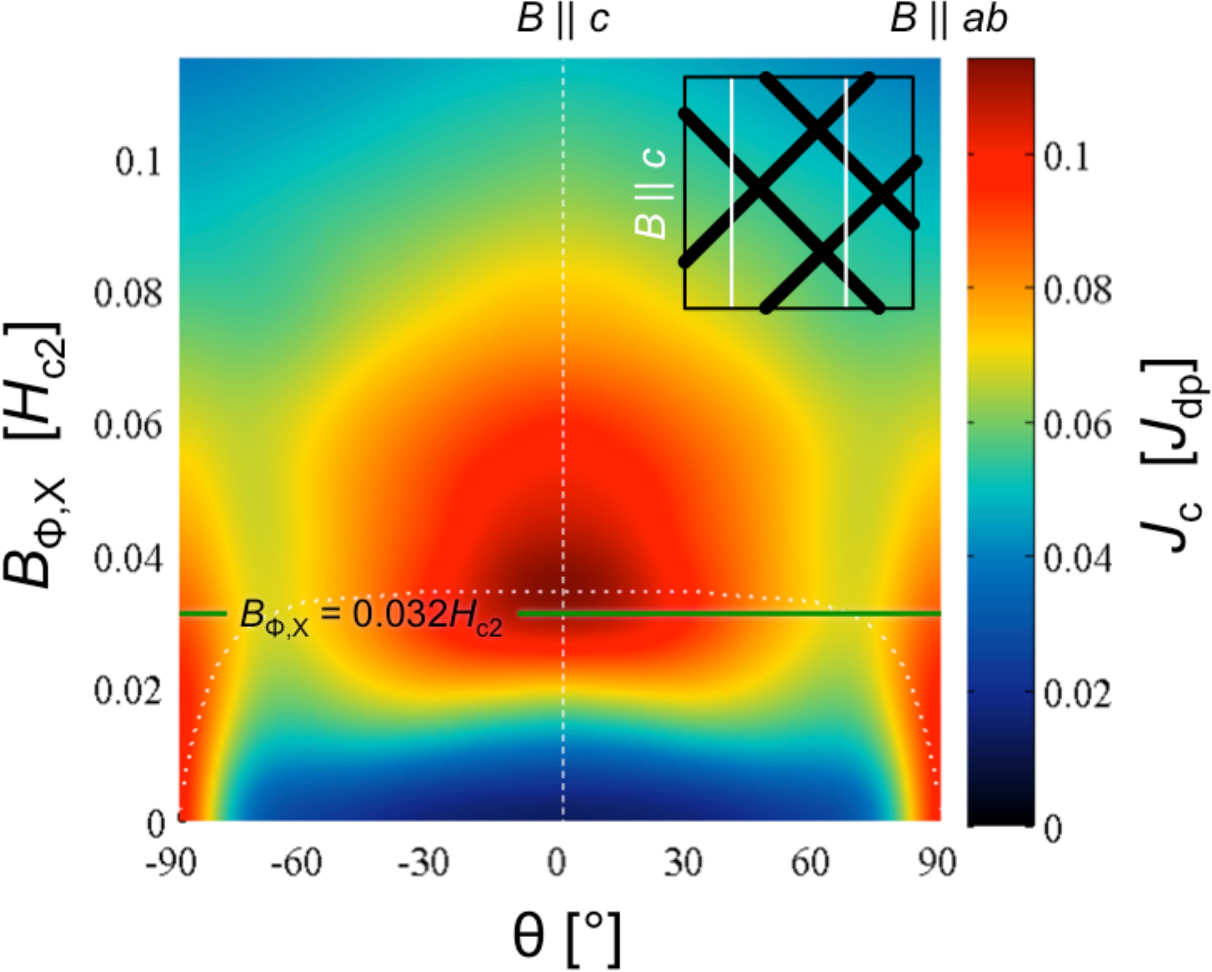}
	\end{center} \vspace{-5mm}
	\caption{
		Angular dependence of the critical current for different concentrations 
		of splayed columnar defects at $\pm 45^\circ$. 
		Similar to Figure~\protect\ref{fig:nanorods_Jc_concentration_theory}, 
		a peak $\Jc = 0.11\Jdp$ occurs at $\theta = 0^\circ$ for $\Bmx = 0.035\Hct$, 
		which corresponds to a volume fraction of 9.3\% occupied by columnar inclusions. 
		The white dotted curve corresponds to the optimal~$\Bmx$ at given~$\theta$.
	}
	\label{fig:splayed_Jc_concentration_theory}
\end{figure}

Non-additive effects can also occur between two identical types of pinning centers. For instance, irradiation induced columnar defects oriented at $\pm 45^\circ$, also known as splayed irradiation, may offer a route towards a marked reduction of the anisotropy of $\Jc$ at an overall high value in single crystals and thin films.\cite{Hwa:1993} The REBCO sample without BZO nanorods was irradiated with 1.4\,GeV $^{208}$Pb ions at both $-45^\circ$ and $+45^\circ$ from the crystalline $c$-axis with a total dose of $\Bmx = 2$\,T ($\Bmc = 1$\,T in each direction). As expected, the Pb-ion irradiation generates perpendicularly crossed damage tracks at $\pm 45^\circ$ from the $c$-axis. The morphologies of the crossed columnar defects are presented in the Z-contrast scanning transmission electron microscopy images of Figures \ref{fig:splayed_TEM_50nm}--\ref{fig:splayed_TEM_2nm_top}. Similar to the REBCO films reported earlier,\cite{Xu:2010} these films also contain a large number of correlated nanoplates and stacking faults parallel to the film plane. We apply the current along the $ab$-plane perpendicular to the splay and applied magnetic field. The changes induced by the crossed columnar defects on the angular dependence of $\Jc$ are shown in Figure~\ref{fig:splayed_Jc_experiment}. We find that $\Jc$ is enhanced over a wide angular range. For the irradiated sample, the increase of $\Jc$ reaches 27\% for $B||ab$ and 71\% for $B||c$. As a result, the critical-current anisotropy is markedly reduced down to 1.17 when defined as $\Jc(B||ab)/\Jc(B||c)$, or 1.3 when more properly defined as $\max_\theta\{\Jc(\theta)\} / \min_\theta\{\Jc(\theta)\}$.

We carried out TDGL simulation of the vortex dynamics in REBCO containing splayed columnar defects. Figure~\ref{fig:splayed_order_parameter} shows a representative snapshot of the arrangement of vortices and defects. The results of the simulation for different concentrations of crossed columnar defects are shown in Figure~\ref{fig:splayed_Jc_theory}. The simulation qualitatively reproduces the reduction of the anisotropy and the overall increase of the critical current observed experimentally in Figure~\ref{fig:splayed_Jc_experiment}. At low concentration of splayed defects ($\Bmx \approx 0.008\Hct$), the simulation of the angular dependence of $\Jc$ reveals two maxima located at $\pm 45^\circ$ as might be expected for a simple addition of pinning effects. However, with increasing concentration the two maxima merge to form a single flat maximum around $\theta = 0^\circ$, as in the experiment. Thus our simulations suggest that a fairly isotropic $\Jc$ can be achieved for relatively low defect concentration with a rather modest increase in $\Jc$, whereas the maximum enhancement of $\Jc(B||c)$ occurs at significantly higher defect concentration accompanied by a sizable reverse anisotropy, i.e. $\Jc(B||c) > \Jc(B||ab)$. 

Figure~\ref{fig:splayed_Jc_concentration_theory} shows the simulated plot of $\Jc$ for different concentrations of splayed defects and angles of applied field. Again, as in Figure~\ref{fig:nanorods_irradiated_Jc_concentration_theory}, one can observe two peaks at $\theta = \pm 90^\circ$ due to intrinsic pinning. The combination of columnar defects tilted at $\pm 45^\circ$ results in a broader peak near $\theta = 0^\circ$ compared to the peak in the angular dependence of Figure~\ref{fig:nanorods_Jc_concentration_theory} for nanorods. The maximum of the critical current occurs at a matching field of $\Bmx = 0.035\Hct$, corresponding to a volume fraction of 9.3\%.

The optimal concentration of nanorods predicted for a REBCO conductor (Figure~\ref{fig:nanorods_Jc_concentration_theory}) is approximately three to four times larger than the values observed in our experiments on currently available industrial samples containing 7.5~\mbox{mol-\%} Zr. Extensive recent work on research-scale samples has shown that the critical current density increases strongly with Zr-content, that is, with the concentration of nanorods,\cite{Selvamanickam:2015,Xu:2014,Abraimov:2015,Selvamanickam:2013,Selvamanickam:2015b,Selvamanickam:2015c} and samples with Zr-doping up to 25~\mbox{mol-\%} have been synthesized. Our simulations (Figure~\ref{fig:nanorods_Jc_concentration_theory}) are aligned with these studies and suggest that there is still room for further enhancement. In contrast, the predicted optimal concentration of irradiation-induced splayed tracks in a pristine sample (Figure~\ref{fig:splayed_Jc_concentration_theory}) corresponds closely to the concentration used in our experiments, and no further improvement is expected. However, we expect that the addition of splayed columnar defects will decrease the critical-current anisotropy of the sample.

Our large-scale TDGL simulations describing the joint action of different dominant pinning centers provide valuable insights for the design of effective mixed pinning landscapes by elucidating the vortex dynamics responsible for the enhancement of $\Jc$ in commercial high temperature superconductors. Visualizing the real dynamics is difficult, as few, if any, experimental techniques can resolve both the microscopic magnetic structure and short timescale. It is however key to achieving higher and more isotropic critical currents in these high temperature superconductors.

\section{Conclusions}

In summary, we introduced the new critical-current-by-design paradigm. This paradigm aims at predicting the optimal pinning landscapes for maximum critical current in targeted high-temperature superconducting applications. We illustrated this concept on technologically important rare earth barium copper oxide coated conductors and validated the results on samples with clearly identifiable pinning effects due to linear correlated defects in the form of self-assembled barium zirconate nanorods and irradiation tracks introduced by heavy-ion irradiation. The TDGL simulations elucidate the vortex dynamics responsible for the non-additive behavior of the vortex-defect interaction in the presence of different types of correlated defects with various spatial orientations. In particular, we observed a vortex-sliding scenario, directly demonstrating the dynamics that can result from synergies between different types of pinning. Using this approach, we also can predict the optimal concentrations of these defects for maximal critical current. As characterization and models of realistic pinning centers improve, and as computational performance increases, the quantitative prediction of a superconductor's critical current from its microstructure now seems within reach.

\subsection*{Acknowledgements} 

We thank I. S. Aranson and L. Civale for illuminating discussions. The computational part of our work was performed on the supercomputers Cooley at the LCF at Argonne National Laboratory and GAEA at Northern Illinois University. The computational work was supported by the Scientific Discovery through Advanced Computing (SciDAC) program funded by U.S. Department of Energy, Office of Science, Advanced Scientific Computing Research and Basic Energy Science. The experimental study at ANL was supported by the Center for Emergent Superconductivity, an Energy Frontier Research Center, funded by the U.S. Department of Energy, Office of Science, Office of Basic Energy Sciences. The operation of the ATLAS irradiation facility at ANL was supported by the U.S. Department of Energy, Office of Nuclear Physics. Use of the Center for Nanoscale Materials at ANL was supported by the U. S. Department of Energy, Office of Science, Office of Basic Energy Sciences, under Contract No. DE-AC02-06CH11357. This paper is a joint, equally contributed theory-experiment work; I. A. S., Y. J., and M. L. have contributed equally to this work.

\bibliographystyle{apsrev4-1-titles}
\bibliography{jc-by-design}

\appendix

\section{Numerical simulations} \label{sec:numerical_simulations}

\subsection{Geometry}

\begin{figure*}
	\begin{center}
		\subfloat{\includegraphics[width=12.6cm]{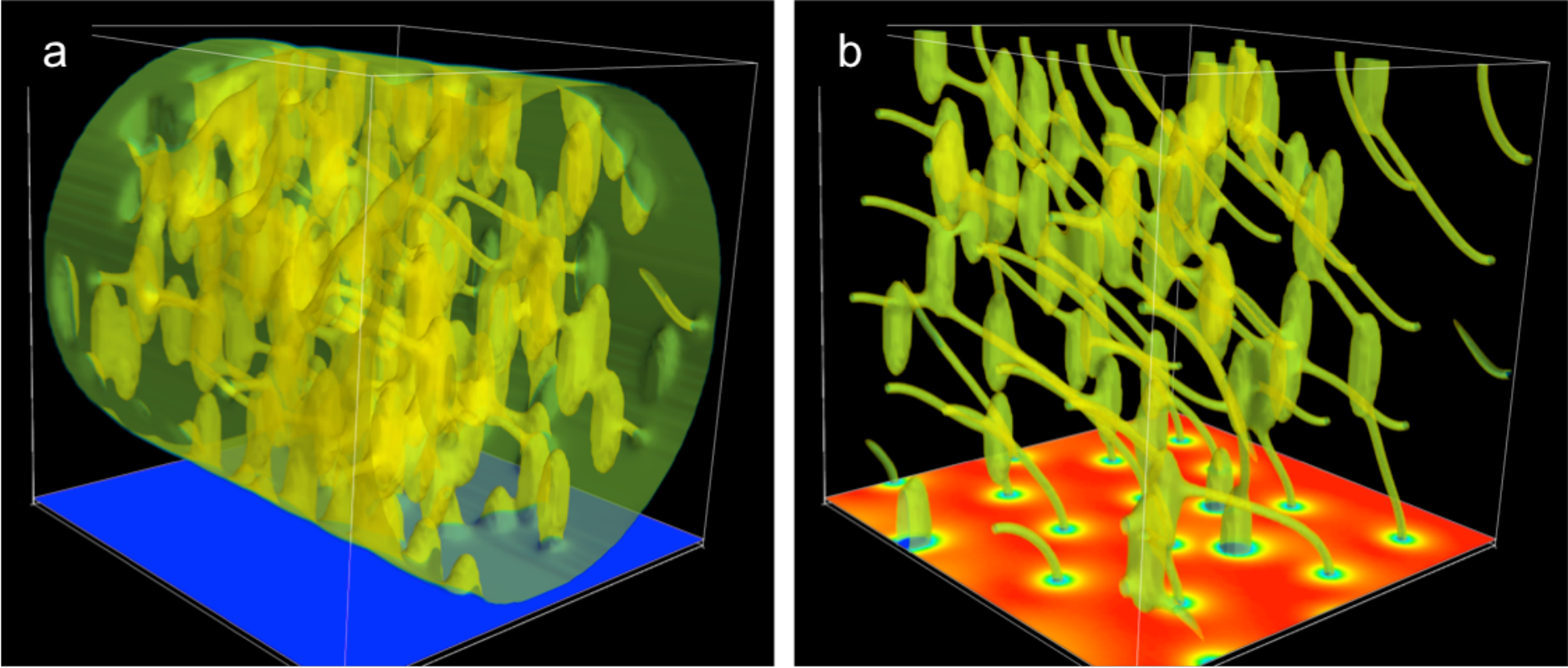} \label{fig:GL_geometry_circular}}
		\subfloat{\label{fig:GL_geometry_periodic}}
	\end{center} \vspace{-5mm}
	\caption{
		Reduction of the geometrical factors affecting angular dependence 
		of the critical current. 
		\protect\subref{fig:GL_geometry_circular}.~Superconducting volume 
		in a cylindrical domain and surrounded by a non-superconducting matrix. 
		\protect\subref{fig:GL_geometry_periodic}.~(Quasi)periodic 
		boundary conditions for the order parameter.
	}
	\label{fig:GL_geometry}
\end{figure*}

We solve the time-dependent Ginzburg-Landau (TDGL) equation for a complex order parameter in a rectangular superconductor of finite size $L_x \times L_y \times L_c = 64\xi \times 256\xi \times 256\xi = 218 \times 870 \times 870$\,nm$^3$ with an underlying periodic numerical grid of size $N_x \times N_y \times N_c = 128 \times 512 \times 768$, where $\xi$ is the superconducting coherence length. We used a mass anisotropy factor of $\gamma = 5$ for YBCO. To simulate the experimental angular dependence, we apply a uniform external magnetic field $B = 0.02\Hct$ and rotate it in the $yc$-plane, see inset in Figure~\ref{fig:nanorods_irradiated_Jc_experiment}. 

In order to minimize surface effects and reduce the geometrical factor in the angular dependence of the critical current $\Jc(\theta)$ we compared two approaches. In the first approach we placed the whole superconducting volume in a cylindrical domain surrounded by a non-superconducting metal and applied an external current along the cylinder, see Figure~\ref{fig:GL_geometry_circular}. This approach shows a uniform angular dependence of $\Jc(\theta)$ for an isotropic system (i.e. with anisotropy $\gamma = 1$) with randomly distributed spherical inclusions. Conversely, a rectangular system with open (no-current) boundary conditions gives 10--30\% variations in $\Jc(\theta)$ under the same conditions. Using periodic boundary conditions in all directions (as shown in Figure~\ref{fig:GL_geometry_periodic}) also gives an almost uniform $\Jc(\theta)$ for the used benchmark system. In that case it is important to use a sufficiently large system such that finite size effects get minimized, i.e., artificial self-pinning by vortices `wrapping' around the system. We choose this second approach because it has 20\% more superconducting volume and thus allows acquiring more statistics of the vortex dynamics.

\subsection{Grid precision}

The TDGL approach can adequately describe the vortex-vortex and vortex-defect interactions on scales comparable or larger than the coherence length $\xi$. Therefore our approach does not require a grid resolution much finer than $\xi$. To determine the required numerical grid precision, we performed a series of simulations on a (32$\xi$)$^3$ sample and varied grid precision $h = L/N$ (distance between grid points) from 0.1 to 1.0. We found that all grids with $h \leqslant 0.6$ give the same $I$-$V$ curves and, correspondingly, the same critical current. We use $h_x = h_y = 0.5\xi$ and $h_c = 0.33\xi$ in the present article.

\subsection{Intrinsic pinning of YBCO} 

The intrinsic layered structure of YCBO can be incorporated in the time-dependent Ginzburg-Landau model in two ways: (i)~by spatial $\Tc(\textbf{r})$ modulation of layers parallel to the $ab$-plane and (ii)~by controlling the effective interlayer distance. The first method is characterized by three parameters: thickness of the superconducting layer, thickness of layer with suppressed superconductivity, and the $\Tc$ in the latter. The second method is based on the Lawrence-Doniach model and uses the numerical discreteness of the mesh as natural pinning landscape. In this model the effective interlayer coupling is controlled by the parameter $\xi / \gamma h_c$, where $\gamma$ is the mass anisotropy factor and $h_c = L_c/N_c$.

We found that these methods typically give similar results that can be mapped to each other for a sample with periodic boundary conditions and without additional inclusions. This means, we can obtain similar angular dependencies $\Jc(\theta)$ of the critical current with a fixed ratio $\Jc(90^\circ) / \Jc(0^\circ)$ by tuning the parameters of the before mentioned methods. However, in reality the intrinsic layered structure has a much smaller scale than the characteristic size of other inclusions. This makes the Lawrence-Doniach model more convenient to describe the intrinsic layered structure of YBCO for large-scale simulations as it, in a sense, integrates out the small scale intrinsic pinning behavior into the parameter $\xi / \gamma h_c$.

Based on a series of simulations of the superconducting sample with given concentration of nanorods and mass anisotropy factor $\gamma = 5$ we found that a Lawrence-Doniach parameter $\xi / \gamma h_c = 0.6$ gives the same ratio $\Jc(90^\circ) / \Jc(0^\circ)$ of the angular dependence as in the experiment (compare Figure~\ref{fig:nanorods_irradiated_Jc_theory} versus the experimental angular dependence in Figure~\ref{fig:nanorods_irradiated_Jc_experiment}). We keep this empirical parameter in all simulations.

\begin{figure*}
	\begin{center}
		\subfloat{\includegraphics[width=12.6cm]{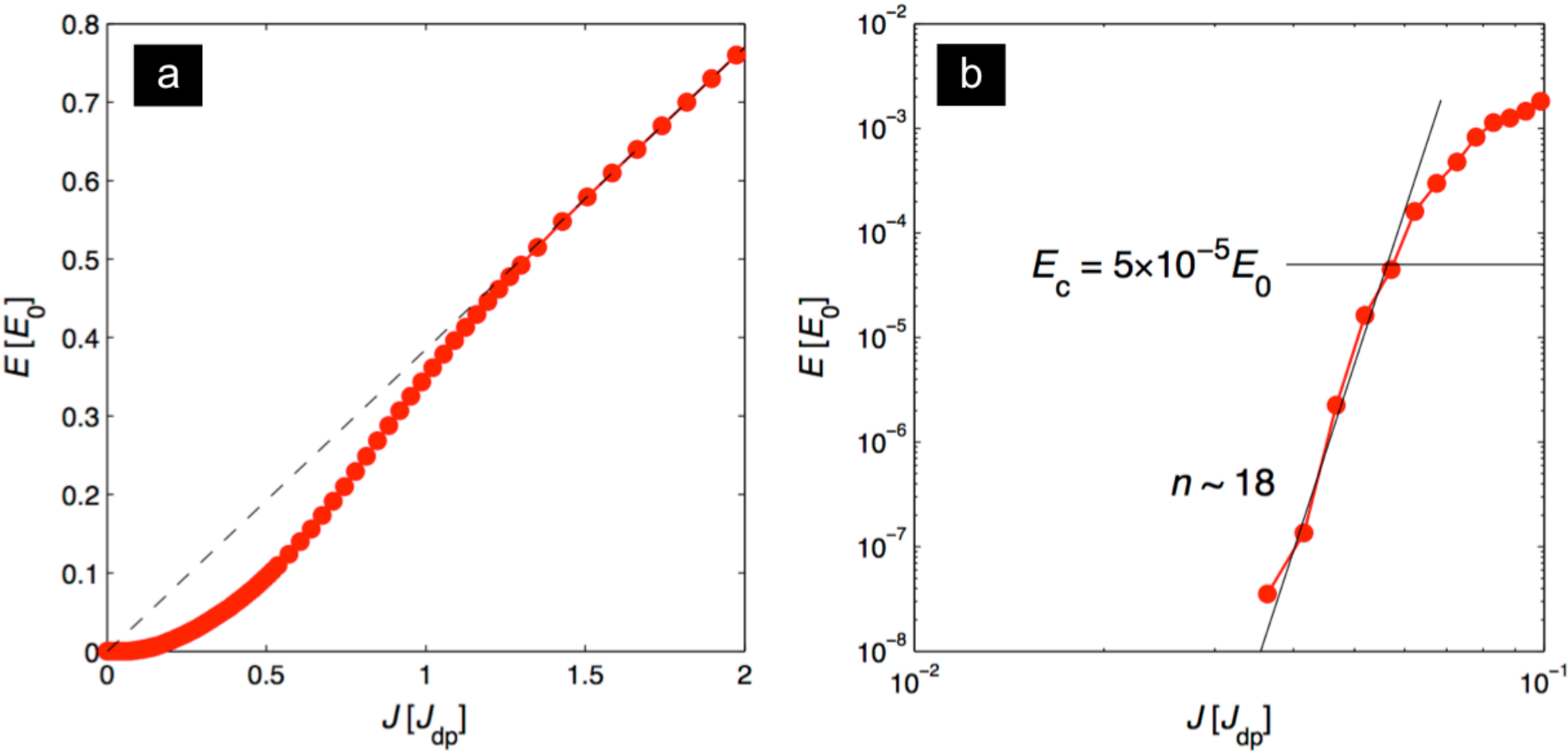} \label{fig:IV_curve_linear}}
		\subfloat{\label{fig:IV_curve_log}}
	\end{center} \vspace{-5mm}
	\caption{
		Example of a simulated $I$-$V$ curve shown in linear scale 
		in panel~\protect\subref{fig:IV_curve_linear} and in double-logarithmic 
		scale in panel~\protect\subref{fig:IV_curve_log}. The critical current 
		is determined by the intersection with a finite electric field level $\Ec$. 
		At currents lower than those shown in panel~\protect\subref{fig:IV_curve_log}, 
		the vortex depinning events are so rare that the calculated electric field 
		becomes very noisy. 
	}
	\label{fig:IV_curve}
\end{figure*}

\subsection{Modeling of inclusions} 

The inclusions are taken into account by the spatial variation of the critical temperature $\Tc(\textbf{r})$, namely $\Tc(\textbf{r}) = \Tcs > T$ in the superconducting matrix and $\Tc(\textbf{r}) = \Tci < T$ inside inclusions. $\Tc$-modulation is naturally embedded into TDGL model by the variations of the coefficient $\varepsilon(\textbf{r})$ of the linear term in the dimensionless TDGL Equation~\eqref{eq:GL}. In this equation we use the coherence length $\xi$ at a given temperature as unit length so that the prefactor $\varepsilon(\textbf{r}) = \es = 1$ describes the superconductor at this temperature. The non-superconducting inclusions with $\Tci < T$ can be described by a negative prefactor, $\varepsilon(\textbf{r}) = \ei = (\Tci - T)/(\Tcs - T) < 0$. Test simulations show that the value of the critical current does not depend on $\ei$ for any $\ei \leqslant -1$. Therefore, we set $\ei = -1$ inside inclusions.

Note, that this approach defines normal-metal inclusions subject to the proximity effect. Although real defects are believed to be insulating in their central region, this still gives a qualitatively correct result as the superconducting material in the vicinity of defects gets distorted as well and therefore suppresses the order parameter gradually.

The prefactor $\varepsilon(\textbf{r})$ and the Lawrence-Doniach parameter are the material parameters of our model. Therefore, the TDGL model can predict the optimal shape and concentration of the inclusions, as long as $\varepsilon(\textbf{r})$ [or $\Tc(\textbf{r})$] describe the inclusion materials realistically.

\subsection{Sizes of inclusions} 

We determined the nanorods' diameter from TEM images of the first sample before irradiation to be about $4\xi$ (with $\xi \approx 3.4$\,nm) and used nanorod dimensions $4\xi \times 4\xi \times 128\xi$ with an average spacing of $16\xi$, which corresponds to a matching field of $\Bmn \approx 0.015\Hct$ ($\Bmn/B \approx 0.75$) and volume fraction occupied by nanorods $\fN \approx 0.030$. The angular dependence $\Jc(\theta)$, shown by the black line in Figure~\ref{fig:nanorods_irradiated_Jc_theory}, does not depend on the nanorods size along the $c$-axis provided they are longer than $\sim 4 \gamma \xi$. Based on STEM images of the irradiated samples, the irradiation induced columnar defects have approximately the same diameter of $4\xi$. The angular dependence $\Jc(\theta)$ shown by the green dashed line in Figure~\ref{fig:nanorods_irradiated_Jc_theory} was obtained for columnar defects inclined at $45^\circ$ with a matching field of $\Bmc \approx 0.032\Hct$ ($\Bmc/B \approx 1.6$) and a volume fraction occupied by the columnar defects of $\fC \approx 0.083$. In order to simulate the plate-like defects around each columnar defects as observed by TEM we added flattened ellipsoids of size $12\xi \times 12\xi \times \xi$ around the tilted columnar defects. The volume fraction $\fP \approx 0.017$ of such defects reduces the peak near $45^\circ$ and increases the anisotropy at $\pm 90^\circ$ as shown by the red line in Figure~\ref{fig:nanorods_irradiated_Jc_theory}.

The angular dependencies shown in Figure~\ref{fig:splayed_Jc_theory} were obtained for matching fields from $\Bmx \approx 0.008\Hct$ ($\Bmx/B \approx 0.4$ and volume fraction $\fX \approx 0.021$) to $\approx 0.032\Hct$ ($\Bmx/B \approx 1.6$ and $\fX \approx 0.084$).

Note, that the simulations do not account for the domain structure of YBCO tape, such as twin boundaries.

\subsection{Finite voltage criterion}

The critical current is determined from the $I$-$V$ curves by a constant electric field criterion with a threshold field of $\Ec = 5 \times 10^{-5} E_0$, where the electrical field unit $E_0 = \rho_{ab} J_0 = (3\sqrt{3}/2) \rho_{ab} \Jdp \approx 2.6 \rho_{ab} \Jdp$ naturally results from the TDGL formalism, see details in Ref.~\onlinecite{Sadovskyy:2015a}. $I$-$V$ curves are obtained by incrementing the external current from zero, while measuring the time average of the voltage across the system at each step, see Figure~\ref{fig:IV_curve}. Since vortex dynamics in the presence of pinning is hysteretic, the $I$-$V$ curve can depend on the external current ramping protocol and the averaging time. To ensure the validity of our procedure, i.e., calculation a unique $I$-$V$ curve, we compared three different ways of obtaining it: ramping the current up, ramping the current down, and fixed current equilibration. All three approaches result in the same $\Jc$.

For YBCO the in-plane resistivity $\rho_{ab} = 5 \times 10^{-5}\,\Omega\,\mathrm{cm}$ near the critical temperature and the depairing current $\Jdp \approx 44$\,MA/cm$^2$ criterion can be estimated resulting in an electrical field threshold value of $\Ec \approx 0.3$\,V/cm. To simulate dissipation levels as low as the industry standard of 1\,$\mu$V/cm, which corresponds to a very small hoping rate of vortices, one would need to simulate the system for a very long time, which is currently beyond reach in terms of computational cost. Typically, an experimental value for the voltage or resistivity is measured during 1\,s. In the simulations we use $10^5$--$10^7$ simulation time steps as averaging time, where each time step corresponds to 0.1 of the Ginzburg-Landau time and the Ginzburg-Landau time is of order $10^{-11}$\,s. This results in a reasonable numerical threshold value for $\Ec$, which is still approximately five orders of magnitude higher than the industry standard, but since the $I$-$V$ curve near the critical current has a sharp behavior (see Figure~\ref{fig:IV_curve_log}), this does not translate into orders of magnitude differences in terms of~$\Jc$. In order to quantitatively characterize this threshold we roughly approximate simulated $I$-$V$ curve near the critical current by a power law $E \sim J^n$ and estimate exponent $n \sim 18$, which means that the higher $\Ec$ criterion yields a critical current $\sim 90$\% higher than the value one could expect by using the industry standard. Thus, our criterion may overestimate $\Jc$ by a factor of two. Such an agreement is remarkable in the field of vortex pinning.

\subsection{Langevin noise} 

The simulation of realistic temperatures used in high-temperature superconductor experiments is a challenge. The absolute temperature in the TDGL formalism is determined by the Langevin noise term in Equation~\eqref{eq:GL}. In Ref.~\onlinecite{Sadovskyy:2015a} the concrete relation between temperature and noise characteristics is derived in detail. In this work we are limited to uncorrelated noise, which results in a technical upper limit in the temperature, which we can simulate-- determined by the numerical stability of the integration scheme. To overcome this limitation, time and space correlations of the noise term are required. However, this is beyond the scope of this work and requires separate investigations. One can expect that a higher temperature leads to systematically lower critical currents, such that the qualitative behavior of our results is still valid (at experimental temperatures used here, the system is still not in the vortex liquid phase).

\section{Samples and measurement methods}

The two types of YBCO coated conductors used in this study (with and without BZO nanorods) were both grown using MOCVD\cite{Selvamanickam:2009} and have a critical temperature $\Tc \approx 92$\,K, coherence length $\xi$(77\,K$) \approx 3.4$\,nm, London penetration depth $\lambda($77\,K$) \approx 260$\,nm, and upper critical field $\Hct($77\,K$) \approx 28$\,T. The critical current densities at 77\,K and 0\,T of the two types of YBCO films are $\approx 410$\,A/cm-width, corresponding to $\Jc$(77\,K$) \approx 4.1$\,MA/cm$^2$, while the depairing current density is estimated to be $\Jdp($77\,K$) \approx 44$\,MA/cm$^2$. The critical current measurements were performed on four-contact bridges with a width of 85\,$\mu$m, a thickness of 1.2\,$\mu$m, and a length of 4\,mm, patterned using optical lithography. The voltage-current characteristic was measured using a four-probe pulsed $I$-$V$ method. The duration of the current pulse was 25\,ms to minimize self-heating effects, and $\Jc$ was determined using the standard criterion of 1\,$\mu$V/cm. The characterization of the angular dependence of $\Jc$ was carried out at 77\,K using a $^4$He cryostat, and the external magnetic field of 1\,T was generated by a triple-axis magnet system which can rotate the direction of the field from $-100^\circ$ to $+100^\circ$ with respect to the film normal.

\section{Heavy-ion irradiation process} 

YBCO bridges were irradiated at the ATLAS facility at Argonne National Laboratory with 1.4\,GeV $^{208}$Pb ions to dose matching fields of 2\,T, either at $45^\circ$ or at $\pm 45^\circ$ from the normal to the film. By using a low enough Pb ion beam current, heating effects are minimal and the transition temperatures of all the YBCO samples remain nearly the same after irradiation.

\end{document}